# Awareness requirement and performance management for adaptive systems: a survey

**Tarik A. Rashid1 · Bryar A. Hassan2,11 · Abeer Alsadoon3,4,5,6 · Shko Qader7,12 · S. Vimal8 · Amit Chhabra9 · Zaher Mundher Yaseen10**

1 Computer Science and Engineering Department, School of Science and Engineering, University of Kurdistan Hewler, Erbil, Kurdistan Region, Iraq
2 Department of Information Technology, Kurdistan Institution for Strategic Studies and Scientifc Research, Sulaymaniyah 46001, KRI, Iraq
3 School of Computing and Mathematics, Charles Sturt University, Sydney, Australia
4 School of Computing Engineering and Mathematics, Western Sydney University, Sydney City Campus, Sydney, Australia
5 Information Technology Department, Asia Pacifc International College (APIC), Sydney, Australia
6 Information Technology Department, Kent Institute Australia, Sydney, Australia
7 Department of Information Technology, University College of Goizha, Sulaimani, Kurdistan Region, Iraq
8 Data Analytics Lab, Department of Artifcial Intelligence and Data Science, Ramco Institute of Technology, North Venganallur Village, Rajapalayam, Virudhunagar District, 626 117, Tamilnadu, India
9 Department of Computer Engineering and Technology, Guru Nanak Dev University, Amritsar, India
10 Civil and Environmental Engineering Department, King Fahd University of Petroleum & Minerals, 31621 Dhahran, Saudi Arabia 11 Department of Computer Science, College of Science, Charmo University, Chamchamal, 46023 KRI, Iraq
12 Department of Information Technology, Computer Science Institute, Sulaimani Polytechnic University, Sulaymaniyah 46001, KRI, Iraq
Email (Corresponding): tarik.ahmed@ukh.edu.krd; bryar.hassan@kissr.edu.krd

## Abstract

Self-adaptive software can assess and modify its behavior when the assessment indicates that the program is not performing as intended or when improved functionality or performance is available. Since the mid-1960s, the subject of system adaptivity has been extensively researched, and during the last decade, many application areas and technologies involving self-adaptation have gained prominence. All of these efforts have in common the introduction of self-adaptability through software. Thus, it is essential to investigate systematic software engineering methods to create self-adaptive systems that may be used across different domains. The primary objective of this research is to summarize current advances in awareness requirements for adaptive strategies based on an examination of state-of-the-art methods described in the literature. This paper presents a review of self-adaptive systems in the context of requirement awareness and summarizes the most common methodologies applied. At first glance, it gives a review of the previous surveys and works about self-adaptive systems. Afterward, it classifies the current self-adaptive systems based on six criteria. Then, it presents and evaluates the most common self-adaptive approaches. Lastly, an evaluation among the self-adaptive models is conducted based on four concepts (requirements description, monitoring, relationship, dependency/impact, and tools).

## 1. Introduction

Self-management can be regarded as one of the essential requirements of a software-intensive system through continuous adaptation to address the rapid changes during runtimes, such as changes in customer needs, changeable resources, and system faults or problems to provide a reliable and robust system [1]. This system must configure, reconfigure, continuously readjust, optimize, and recover itself while keeping its complexity hidden from the user [2]. Self-adaptive and self-managing methods have been used in various application areas like control systems, programming languages, fault tolerance, robotics, control systems, and many other application areas [3].

It would be difficult to use, understand, and even apply for an application when software changes. In this situation, the most suitable solution is re-engineering. Within the re-engineering process used to repair and reform the system, the organization needs to evaluate how the high-level requirements are applied to the design and how modifications are incorporated to comply with the required changes. This re-engineering process may require a form of reverse engineering to obtain an abstract representation of the software [4]. Awareness requirements are non-functional requirements that represent both the environment of the system and the software itself to be able to adapt.

The demand for adaptive software systems capable of adapting to the changes within the environment and being aware of their resources is growing dramatically as users' needs are constantly increasing over time. Such systems would consist of specific functionality that requires feedback loops to keep monitoring the execution processes and the adaptivity operations. Autonomous and automatic systems usually fall in this category of adaptivity since almost all the electronic machines, equipment, and Embedded systems uses such techniques continuously and depend on their automation, especially in real life-critical situations that might be needed 24 hours a day, which requires continuous updates and requirement or else a failure would lead to a catastrophic result. In essence, this is the primary goal of the adaptive systems, to keep up with the dynamic changes in performance, functionality, and operation [5].

Feedback loops are necessary in this case as they provide a monitoring mechanism for adaptive systems as well as helping in the process of analyzing the systems and their functionalities. To achieve an acceptable analysis, the requirement of the run time, success, failure, quality of services, and others must be defined [5].

This paper reviews requirement awareness and self-adaptive systems, their characteristics, and the procedures taken to engineer the self-adaptive system. The paper also explores a few research areas related to the subject. In the first Section, a quick overview explains the adaptive system and its relation to requirement awareness. Some of the problems that requirement faces in a general view and how to solve them are mentioned in Section 2. In Section 3, the most recent researches done about this subject will be explored. Moreover, the classification of the self-adaptive system is provided in Section 4. Finally, a brief conclusion is given.

## 1.1. Adaptive Systems

An adaptive system or complex adaptive system (CAS) is the system that can change itself in reflection to changes within the environment in which the system is operating. The adaptive change that the system is willing to achieve is generally related to achieving the goal or objective to which the system was initially designed. Generally, adaptivity is related to animals and plants as they tend to adapt themselves to the changing environment. However, now simple systems can also be adaptive as they adapt to their environment. Although these environmental changes cannot be easily understood just by considering the impact they propose, indirect effects should also be considered, which, in turn, causes the adaptive response. Understanding how indirect effects arise from adaptation can aid in recognizing how to affect the system in the desired way.

The system's capability to change its behavior as a reaction to its surroundings, operational context, or environment is commonly known as Self-Adaptivity [6]. The "Self" part means that the system can decide autonomously or on its own (with no or minimal interference) on adapting or changing itself to fit the changes in its environment. Although many self-adaptive systems work without requiring any human interaction, some of them need some guidance in the form of high-level objectives.

As previously mentioned in the introduction section, the users' needs and dependability on such systems are increasing dramatically. Thus, the complexity of these systems will increase, and it will be more challenging to manage their automation, reliability, and validity. So, the design phase must include the implementation of these systems flexibly.

In the entire self-adaptive systems development and at each phase, giving attention to quality assurance is essential. This begins with collecting the requirements, architecture design of the system, and lastly, the execution, testing, and system deployment. In each phase of the evolution, the quality of the artifacts influences the other parts of the system since all the details are correlated to each other strictly [7].

The definitions mentioned above generally define runtime changes connected with the system's functional and non-functional requirements. Although many software engineering researches were centered only around the non-functional part of the system. [8] divided the mechanisms of adaptation into few processes: monitoring software entities (self-awareness) and the environment (context-awareness), analyzing significant changes, planning how to react, and executing, where decisions can be made. Generally, an external adaptation manager holds the adaptation process, which is separated from the application logic [9].

Furthermore, the Proposed MetaModel presented by [10] aims to classify the adaptive systems into four types based on their requirement and implementation. The types are:

1. Adaptation of Type I is regarded as the most straightforward implementation of intelligent systems.
2. Adaptation of Type II is composed of complex systems.
3. Adaptation of Type III illustrates an advanced implementation of a smart system.
4. Adaptation of Type IV is considered the higher level of a smart system.

## 2.2. Awareness Requirements and Adaptive Systems

Changes within the system's environment, operational contexts, along requirements lead to an adaptation process by that system. When software systems (self-adaptive software systems) change, this will lead to difficulties regarding understanding, using, and maintaining the system. The undesirable effects of such changes get even worse since the team applying those changes is mostly not the team that developed it. After few iterations of changes to the system, it will gradually become far away from its requirements. Leite claims that re-engineering is a well-suited approach for a case where maintenance is independent of the process through which the artifact is created [11]. He also defined the re-engineering process as the examination and alteration of a subject system to reshape it into a new form along with the succeeding implementation of the new constitution.

Re-engineering a system usually requires reverse engineering to achieve a more abstract description of the software, then some reconstruction or forward engineering is followed. Many researchers believe that self-adaptive systems implement awareness requirements, a higher-level abstraction for a self-adaptive system. Awareness requirements may be defined as non-functional requirements associated with the system's ability to perceive or be aware of its surroundings. In the goal-oriented model, awareness requirements, like non-functional requirements, are typically listed as a soft goal. The non-functional requirements affect the decision and implementation of software development and maintenance associated with quality attributes. The development of a new self-adaptive system and changing the non-adaptive into an adaptive approach is considered a big challenge by many researchers in software engineering.

Previously conducted studies have been based solely on using a single system and have not considered the set of issues that might emerge between distinct systems [12]. It is known that awareness requirement is the type of study whereby the steps aimed to define requirements linked to the adaptability of a self-adaptive system [12]. It illustrates a technique in which it sets out the problems to be worked out through self-adaptation and the scope needed to cover the problem areas. For this to be done, both the pattern and the type of the requirements are defined to set up the Goal-Model [3].

Although architects with experience can help engineers to recognize the awareness requirements from the information or knowledge available in the source code of the system, such as goals and soft goals, functional and non-functional requirements, architectural decisions, design choices, and implemented solutions, having an outdated requirement or poor design will create a barrier for this task. To obtain improved results, a systematic approach could be used for this purpose.

## 2. Previous Literature Works

The increase in the usage of adaptive systems led to the rise in the research area for defining requirements for such systems. Therefore, many approaches and methodologies have been introducing from a different points of view. The main focus of these researches was to establish a common ground for defining comprehensive requirements for self-adaptive systems. The literature review section offers a review of the most recent studies conducted in this field. In the surveys conducted by [13] and [4], taxonomy has been introduced to overview the approach researchers follow to analyze requirements for self-adaptive systems. The survey results exhibited that the majority of the methods the researchers were mainly concerned with the control side of the problem, including the mode, type, loop dimensions, and schemes. This shows that the focus area of current research is done to demonstrate how the requirements can be controlled through the suggestion and analysis of models based on their impact. The second category of researchers focused on analyzing an existing target system that included application domain, performance domain, and dimensions. The main scope of this group is to define the requirement based on the target domain. Firstly, the domain and performance details are analyzed, and then the conditions are set for them afterward. The minor focus that the taxonomy showed was the requirement validation. Case studies and simulations are considered within the validity, indicating that the requirements are analyzed based on the previously introduced case studies and the simulation process conducted to verify its validity and as shown in Figure (1).

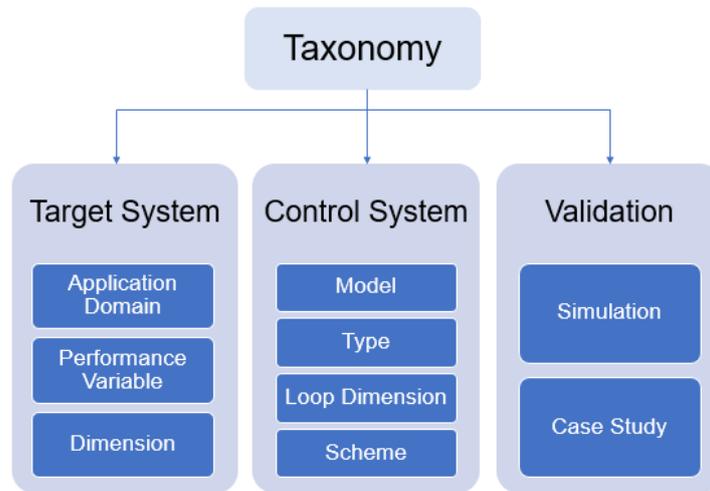

**Figure (1): Design of self-adaptive software systems with the use of control engineering approaches**

Another study presented a systematic literature review approach for modeling the requirements for self-adaptive systems [14]. This system is depicted in Figure (2). The work was mainly focused on dividing the requirement analysis stages into several distinct phases. These phases included planning, conducting, and reporting. By having these phases, the process for gathering requirements was divided into smaller and more controllable pieces. The first phase is planning, in which all researches are conducted on the target system to explore its needs and ask further questions that lead to the desired results. Furthermore, a protocol is also developed during this stage in which it includes every detail about the possible requirements, criteria. Strategies and quality assessments. The next stage is the conducting stage, which involves identifying the most relevant requirements and needs and when to adopt them. All the decisions and negotiations are made in this phase, giving a result output that the system will be designed on. At the end of the conducting phase, all the decisions are documented in the reporting stage, finalizing the gathered requirements and preparing them to transition to the design phase.

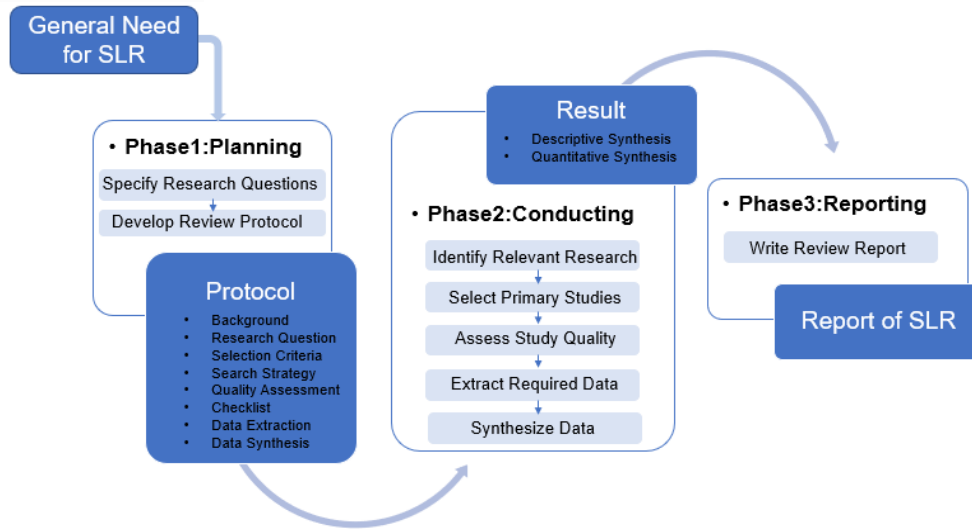

**Figure (2): Requirements modeling and analysis for self-adaptive systems.**

A preference-based reasoning and automated planning to enable continuous adaptive reasoning of requirements at runtime approach were introduced by [15]. This research focused on providing a framework on the behavior of systems at runtime and the techniques used to define the requirements based on predictions. Figure (3) elucidates the reasoning process at runtime.

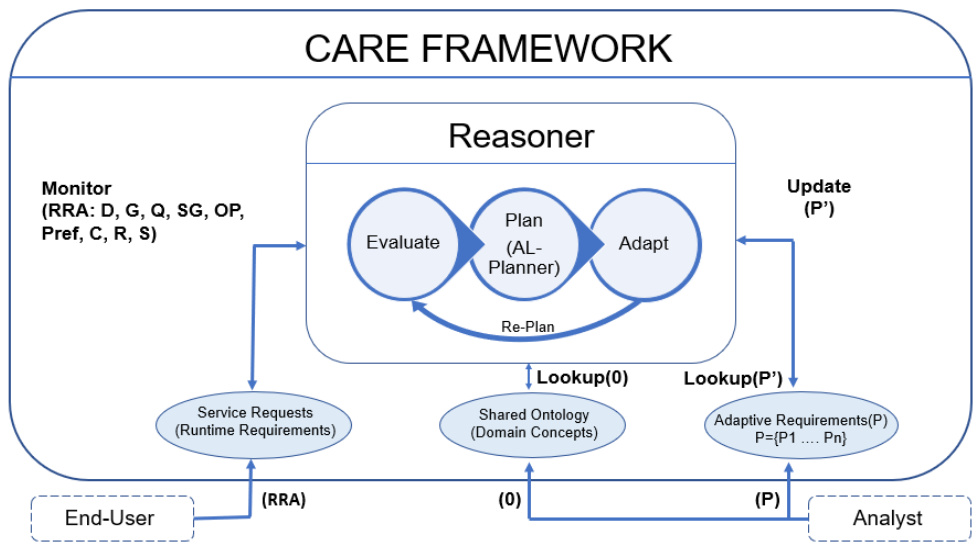

**Figure (1): Reasoning process at runtime.**

Feedback loops designed to control system outputs and adaptivity in case of failure are the core for many previous proposals [16], followed by an approach that questions the actual requirements that lead to feedback loop functionality. Moreover, they have proposed an architecture that presents adaptive systems as a control system reliant on the feedback loop, as illustrated in Figure (4). This process starts at the very late stage of the requirement phase and focuses mainly on the assigned software requirement rather than the fundamental goals and targets behind them. The reference input here is the system requirements which act as a trigger to activate the following processes. The expected output is an indicator of requirements convergence which means the percentage and ability to recognize the functional requirements during the runtime process. Non-functional requirements are also a portion of the convergence in which its boolean nature indicates the degree and rate of satisfaction. The general framework is goal-oriented and predicts the output of vanilla requirement engineering phase ontology. In an attempt to implement the loop, the information of the target system should be logged using instrumentation techniques. The adaptivity framework should manipulate the parameters of the system to complete the target system.

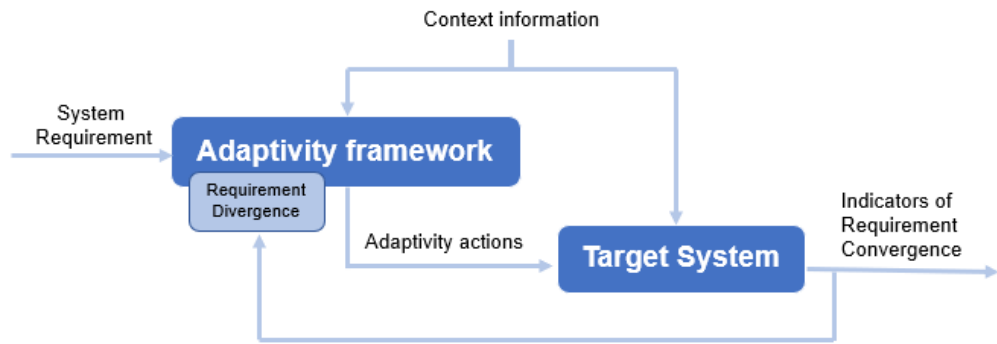

**Figure (2): View of an adaptive system as a control system.**

[17] worked on a comparison between requirement engineering (RE) at the design level and at the runtime. Figure (5) explains the RE at the design phase start at the beginning of the requirement phase with the goals and requirement elicitation. The primary stakeholder goals are analyzed and elaborated to help come up with alternative operationalizations. The requirements at this level are called adaptive requirements, which are divided into functional and non-functional requirements. All the requirements have a group of flexibility and adaptivity in different conditions. These characteristics allow monitoring Specification and evaluation during the early level of RE. The RE at runtime gives the analyzing attributes to the system. The key stakeholders act as direct users at the runtime. It starts with goal-oriented requirements specification as an input that encompasses the planned requirements at the specification level. The new requirements are continuously recorded and expressed as user requests, and they are identified by user itinerary monitoring.

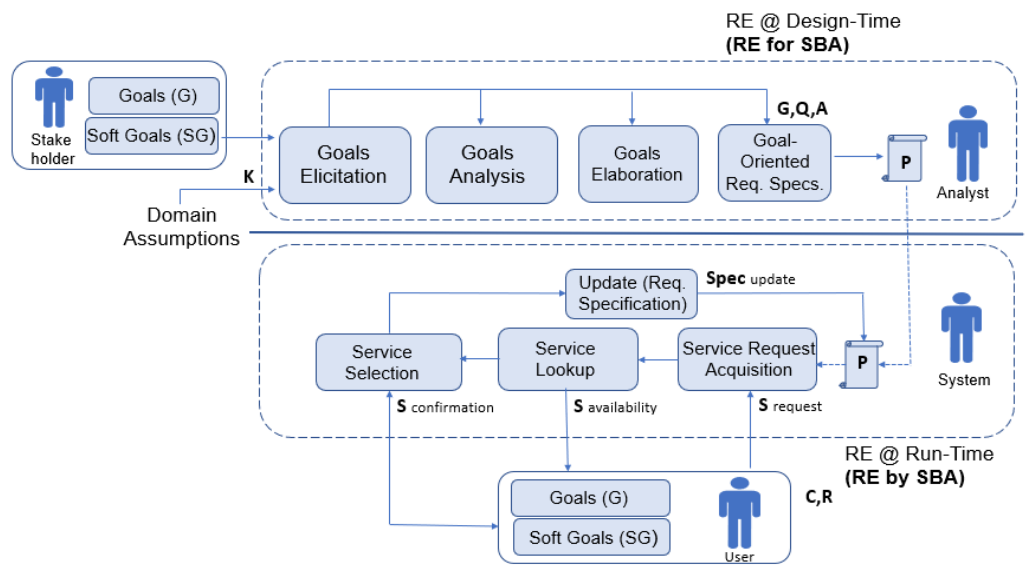

**Figure (3): Requirement engineering process at runtime versus design time.**

[18] created a formula for requirement problem that is locating the Specification (S), which will fulfill the requirement (R) without neglecting the assumption (K) of the relevant domain knowledge, to satisfy the requirements.

$$K, S \vdash R \qquad\qquad\qquad (1)$$

The formula above highlighted the significance of the Specification to be constant with assumptions of the domain, through which both K and S requirements should be measured. It was finally claimed that there is an additional relationship to the problem of requirements than what the formula above has stated [19]. Specifically, a new system in the issue of requirement is used to perceive that similarly to goals and tasks which are suggested from the core of ontology for requirement (CORE), different customers may have unique choices over conditions, that they may be involved in the decision-making between candidate solutions available for the encountered requirements problem, with probably numerous candidate solutions being available such as the ones available in service-/agent -oriented systems, whereby specific service-/agent can also compete by providing the equal capabilities, and not fixing that requirements, but trade with new records from the operational environment or the stakeholders. When there is a lack of preference availability, it can be noted that it is:

- Not understandable as to how to provide a comparison between requirements candidate solutions,
- The type of standards (should) be applied for comparison purposes, and
- How these standards are depicted in the models of requirements.

To sum up, modeling of requirements language turned into a stepping stone for the evolution of the latest modeling of requirements languages that can be employed to show records and carry out reasoning to resolve requirements. The schematic symbol found in Tropos [20], which consists of the most effective minimal principles and relations, is required to formalize the properties of its answers and the problem of requirements. Techne can be regarded as a manageable formalism technique for the runtime requirement issue since it is adjusted to the concept, assumption of task domain, goals, and relations. For example, consequence, Is-optional, Is-mandatory, Preference, persist in applying for the SAS's requirement engineering. To maintain the simplicity of discussion within this article, it is supposed that the set of requirements and the various information available in the proposition are not anything but a natural language sentence. Definitions of consequence relation that define the candidate solution are also provided.

Our review research differs from previous survey work in several areas, which are described below: (1) Previous research is out of date since the most recent study was conducted in 2017. None of them focused on prior review articles or the categorization and assessment of existing self-adaptive systems. This survey covers the most recent developments up to and including 2021. (2) the criteria for awareness and adaptive systems are detailed, including a step-by-step method. (3) Additionally, we address recent reviews on adaptive systems and their variants. (4) A short assessment of the self-adaptive is performed using four concepts (description of needs, monitoring, relationship, dependency/impact, and tools). (5) We found over 100 research articles by doing a Google Scholar search using various variations of the keyword 'self-adaptive systems and awareness requirements.' After a thorough examination of the collected articles, we identified 30 as the most important in self-adaptive systems and their variants discussed in this research.

## 3. Classification of Self-adaptive Systems

The capability of systems to accommodate changes in response to the environment is commonly known as self-adaptivity. The system can decide autonomously how to accommodate variations in their environment or context, hence the prefix of the "self". Even though self-adaptive systems can operate without human intervention, it is helpful to get guidance through policy logic [9]. Depending on the behaviors of adaptive systems, four properties could be considered for self-adaptivity [21], each of which covering a particular set of goals mentioned below:

1. Self-configuring: the ability of automatic reconfiguring and capability to respond to the change dynamically by installation, upgrading, amalgamating,
2. Self-healing: this is often considered as the ability to regulate, analyze, and respond to interruptions,
3. Self-optimizing: which could be the ability to handle performance to satisfy the various requirements of users, such as throughput and response time,
4. Self-protecting: this is the recognition of security attacks and alleviating their effects.

Self-adaptive systems (SASs) are expected to alter their behaviors concerning dynamic and environmental changes to fulfill functional and non-functional requirements [14]. Due to the ongoing communications between the environment and software, SASs have experienced some difficulties in satisfying its criteria on some quality attributes. For instance, fault-tolerance and replaceability, to mention a few. to maintain these attributes, adaptation mechanisms in SASs must be created to provide them with the ability of self-healing, self-configuring, self-optimizing, and self-protecting, which are recognized as self-properties. Consequently, when developing SASs, both domain logic and adaptation logic should be considered [14]. Classification will be thoroughly explained, along with highlighting the importance of awareness requirements. In this Section, the 5W+1H questions will be explored to elicit adaptation requirements [22]. Table 1 shows the classification of self-adaptive systems.

**Table (1): Self adaptive system classification**

| Dimension of the taxonomy | Questions |
|---|---|
| Time | When to adapt? |
| Reason | Why do we have to adapt? |
| Level | Where do we have to implement the change? |
| Technique | What kind of change is needed? |
| N/A (Nature of a SAS leads to an automatic type of adaptation) | Who performs the adaptation? |
| Adaptation Control | How to perform the adaptation? |

Additionally, Figure (6) depicts the design of self-adaptive software systems using control engineering approaches.

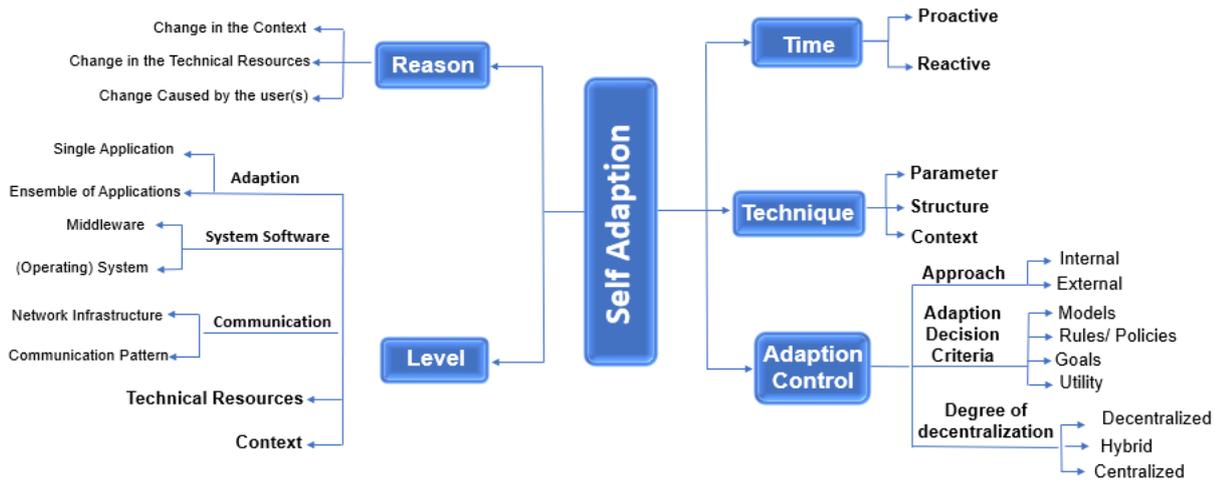

**Figure (6):  Design of self-adaptive software systems using control engineering approaches**

### 3.1. Time

"When to adapt?" is associated with the time aspect. [23], divided the temporal factor into two-dimensional adaptation: (i) reactive and (ii) proactive. To describe this aspect, it is sufficient to segregate between adaptation before and after adaptation. Reactive adaptation is the traditional view, in which it works when an event has occurred where adaptation is needed (like a change in performance or resources). Whereas, in proactive adaptation, the system does not wait for an event to occur for the adaptation to happen. Proactive adaptation is executed before an application asks for adaptation.

"When to adapt?" is a central and essential question to be asked. The algorithms that are predicted and used for adaptation and are proactive have many problems. They are considered difficult to develop and may lead to faulty results causing malicious suboptimal performance. In addition, their appropriateness depends highly on the specific prediction tasks. Nonetheless, proactive adaptation is favorable from a user's point of view, as the interruptions in the user's workflow with the system are avoided. Hence why many approaches are more focused on reactive adaptation. However, this does not change the fact that reactive or proactive adaptation is not exclusive.

The process of adaptation's mapping to the Autonomic Computing MAPE cycle, which is the central feature of the adaptation, is the monitoring of the environment, figuring out adaptation plans, analyzing for change, and the procedures to be executed. Proactive and reactive adaptations differ in the analyzing phase strongly. With a bold adaptation, the data monitoring is used for the systems behavior forecasting or that of the states of the environment. With reactive adaptations, for abnormal patterns, the data monitored but are pretty similar regarding the set of activities on planning, monitoring, and executing can be analyzed [24]. A possibility can be a combination of reactive and proactive adaptation. The adaptation goal can be the bold adaptation. The backup mechanism is the reactive adaptation, i.e., a case where a variation is not predicted (a component failure, per se).

Monitoring type counts as an additional time-related aspect. Adaptive monitoring indicates that a set of selected features are monitored, and if anomalies happen, the monitoring process is intensified. Continuous monitoring defines a monitoring effort that is constant in terms of the environment and resources. The choice between adaptive monitoring vs. continuous tracking depends on the monitoring cost, and in turn, on the process of self-adaptation. However, we cannot disregard that this is not related to the decision of adaptation itself that it is an implementation detail, which is why it is not included in the taxonomy.

## 3.2. Reason

Generally, the process of adaptation is triggered by a change. While the source of the change determines the sufficient adaptation parameters, this reaction is considered costly. To decide whether an adaptation is needed, both the type of the reaction and the impact should be identified clearly. That is why it is of utmost importance to define "Why do we have to adapt?" the reason. This is an essential question affecting the reaction, as different adaptation activities are formed for various reasons.

In a SAS (Self-Adaptive System), the change in either one or several elements of the system is regarded as the reason for the adaptation:

1.  Technical resources change, e.g., a fault in the software, a defect in a component of the hardware, a network connection's alternative availability,
2.  An environment change, e.g., a change context variables state, or,
3.  A change related to the user, e.g., might include the preferences of the users or a variation in the user group's composition.

In this taxonomy, variation in context, technical resources, or the involved users results in adaptation. The user's actual involvement in the reason of adaptation has not been specified in other classifications. (Self-) adaption's taxonomies generally do not mention the reason for adaptation, as adaptation reason is not identified as an element of the decision. Here, we especially mention the adaptation reason because this is the catalyst that triggers adaptation. In our knowledge, a significant part of deciding on the transformation is the adaptation reason. Therefore, adaptation needs to be a part of the self-adaptation features since they influence the monitored elements. The need for adaptation cannot be specified without sufficient process monitoring, resulting in a delay in starting the adaptation process. Subsequently, the monitoring of the technical resources, the software and the hardware, the user's interactions with sensors and interfaces for user interaction, and the environment is essential [22].

## 3.3. Level

Adaptation can be carried out at various levels in the system for finding the answer to the question "Where do we have to implement the change?" which can be referred to as the level of adaptation. First, we need to familiarize ourselves with the levels of a SAS. Generally, a self-adaptive system is composed of distinct elements. Typically, a SAS incorporates the following elements: the adaptation logic and the managed element(s). While the managed elements can be adapted, the adaptation logic often stays stable for the technical resources like the control unit. Furthermore, adjusting the user(s) and the environment causes the common view on a self-adaptive system to extend. The atmosphere is not changed but monitored only. The logic of adaptation facilitates the improvement of performance over time.

The managed elements consist of different levels. The technical resources are usually hardware and other managed resources, such as smartphones, computers, traffic signs, robots, or production facilities. The system software controls the hardware like middleware in a distributed system and an operating system. The application is established atop the system software. This application could be a distributed application divided into the parts of the application running on various devices simultaneously or an application that runs on a single machine. Consequently, these two are both seen as a single application. Moreover, numerous different applications can run simultaneously and collaborate and take all the application parts together. This often leads to interferences happen, i.e., dependencies regarding the use of resources and interceptions that are undesired. Communication is essential for the logic of adaptation elements and the interaction between the managed resources. Communication can be referred to in two ways, where logical communication is the communication pattern, which can be regarded as the interaction style among the elements. While the network infrastructure is the way, the network is physically connected through routers, network cards, vast local area networks, etc. Potential implementation could be pub/sub communication or event-based communication. The system's context is an additional level apart from software and technical resources.

Systems that are context-alert are capable of adapting their context. This results in extending the self-adaptive systems view, where the environment is not precisely modified but is monitored. Essentially, this can be accomplished using technical resources, specifically actuators, but the adaptation logic hasn't explicitly controlled it. Currently, context adaptation is not counted as one of the levels of adaptation for self-adaptive systems.

Adaptation can occur on all levels: an example of adaptation in the application level can be observed when the smartphones are turned to silent mode during a meeting and identified if the meeting is scheduled in the calendar. Components exchange at runtime is offered by adaptive middleware. Adaptation at the network level is possible by the autonomic communication techniques, e.g., from 3G to WLAN as far as a WLAN connection is available. Back-ups of systems can have an automatic start that is enabled through self-healing capabilities. For instance, modifications can be caused in the technical resources in data centers. Context-adaptive applications can adapt the contexts by using actuators or adapting their behavior to the surrounding context. An example of this can appear in smart meeting rooms. When the presentation starts, the light dims automatically. The self-adaptive systems adaptation logic needs to know the difference in levels mentioned, alternatives, possible adaptation, and identify adaptation plans for proper levels. Thus, to accomplish the system's goal, we need to find a correct answer to the "Where to adapt?" question depending on the level of where adaptation is required. In the event where adaptation can affect various levels, the plan to perform based on the criteria like cost, in terms of accomplishable utility, or the time for the system, must be decided by the adaptation logic. User adaptation is possible in theory but not preferable in practice because the application is built for the user, not the other way around. That is why the user adaptation is not included in this taxonomy.

### 3.4. Technique

Certain adaptation action needs to be done on the levels where it is identified that adaptation is necessary. Hence the question "What kind of change is needed?" is explored. A possible answer is a technique. Thus, it is not sufficient for only the levels to be identified for adaptation to take place. Different methods for adaptation can be found in the literature. [25] distinguished between two approaches for adaptive software, which are:

1. Compositional adaptation,
2. Parameter adaptation.

The exchange of system components and algorithms dynamically is enabled by compositional adaptation. This makes it applicable for defective parts and, in turn, leads to increased performance by adding new elements or adjusting the system to new situations. Suppose the example of the server is used. In that case, it can be understood that during the execution of the application, a new instance of the server can overtake the tasks and responsibilities of another example of that server, or even new instances can be integrated into the system.

Altered system behavior is achieved by parameter adaptation through altering the parameters of the system. This can be accomplished quite easily, as the adaptation logic needs only to handle and modify the parameters. However, if the parameters are dependent on each other, the modification of these parameters will be pretty complex. Moreover, it is possible to switch between algorithms for a component that has different algorithms. Furthermore, at runtime, the dynamic integration of new algorithms is not feasible. New elements cannot be integrated dynamically too. A rule-based system can be an example of parameter adaptation. Depending on the servers' workloads, the number of servers is identified based on the rules that depend on the situation. Only the addition of new servers is achievable in this circumstance as there are available servers that are configured. Nonetheless, the dynamic relocation of one instance of a server to another is not possible since it could disrupt the structure of the system.

### 3.5. Adaptation Control

A self-adaptive system incorporates the adaptation logic along with the managed resources, where the adaptation logic is in charge of performing adaptation control and includes the user(s), the environment, the resources collected, the adaptation planning, executing the adaptation plan, and analyzing the anomalies of the data monitored. Hence, the answer to "How is adaptation performed?" is mainly responsible for adaptation logic. By providing answers to such questions, the decision on how the adaptation performs is decided by the adaptation logic. Two methods for the implementation of the adaptation logic are found in recent literature. Self-adaptive systems following the external method are divided into the adaptation logic and the managed resources, which leads to the rise in maintainability done by modularization. The internal approach combines the system resources with the adaptation logic.

A metric is needed for the control unit to choose how to adapt. Various metrics are present in the literature: policies, models, goals, rules, utility functions. The best adaptation possibility should be selected depending on the analysis of all options based on the criteria. Also, it is necessary to mention that different standards can be combined, such as planning of goal model-based. For solving the conflicts between the goals, additional utility is added.

The degree of decentralization can be considered as another characteristic of adaptation logic. The solution for the small number of resources needed to be managed is a centralized adaptation. But for large systems with various components and aims of improving the system performance for adaptation, a decentralized approach is much more preferred. There are many decentralization levels. A fully decentralized method where every sub-system needs to have their logic of adaptation complete in which various communication patterns are possible. If an approach is a hybrid, the sub-systems contain the distributed functionality of adaptation logic, or the decentralized methods contain central components.

## 4. Approaches of Self-adaptive Systems

In the following sub-sections, the popular approaches used for requirement gathering for self-adaptive systems will be illustrated. KAOS (Knowledge Acquisition in automated Specification).

### 4.1. KAOS

Referencing to [26], KAOS is essentially a Goal-Oriented Requirements Engineering (GORE) framework [27] where semi-format and format reasoning of behavioral goals are on high emphasis to goal refinements derivation, risk analysis, management of conflicts, and operationalization. Sub goals in KAOS are usually derived from plans through decomposition. Goals might also be improved into requirements (assigned to software agent) or expectations (transferred to the environment agent). KAOS highlights the obstacle concept to be the case in which the goal achievement is prevented. The solution for this situation can be depicted as a goal which is further refined and improved in the form of requirements and expectations. KAOS specifications can be built from high level goals as follows:

1. Goal development: The goals are refined by detecting additional distinct goals for which the high-level goals can be distinguished.
2. Object identification: object identification of goal formulation specifies the links among them and the domain's properties.
3. Operation identification: defining the state transition of the object that can be considered essential to the goal.
4. Goal operationalization: description of the operations satisfying all goals.
5. Responsibilities assignment: where agents have mapped to the leaf goals and mapping operations assignment to agents.

### 4.2. SysML

SysML is regarded as a general-purpose modeling language intended for system engineering applications. It was developed back in 2001 and was initiated by the Object Management Group (OMG). The International Council on System Engineering (ICOSE) with Conrad Bock, Cris Kobryn, and Sanford Friedenthal were involved [27]. The current version of SysML is V1.5 which was issued in May 2017. A wide range of systems and systems of systems specification, design, analysis, verification, and validation are supported using this modeling language. These systems entail hardware, software, and information. SysML modeling language entails a graphical construct that is used to define the text-based requirements and associate these requirements to respective model elements. System requirements behavior, parametric and structure are modeled by SysML as it provides a graphical representation with the semantic foundation for this purpose. The requirements diagram captures requirement hierarchies and requirements derivation and the <<satisfy>> and <<verify>> relationships allowing a modeler to relate a requirement to a model element with a bridge in between the typical requirements management tools and system models provided through the requirements diagram.

### 4.3. SysML/KAOS

This model is a goal-oriented strategy that helps identify and resolve conflicts between needs and aids in the search for alternatives. The SysML/KAOS language extension to the SysML requirements language incorporates the two most widely recognized and used goal-based methods in requirements engineering over the past decade. The primary objective is to give a notion of aim in SysML, a requirement model, by using SysML's ability to provide a coherent description of requirements and their connections. To accomplish this, a combination of the KAOS concept for expressing functional needs and the NFR model idea for expressing non-functional requirements is the most appropriate concept for specifying non-functional requirements. Non-functional needs will be incorporated before functional requirements and at the same abstraction level as functional requirements. Additionally, this method facilitates identifying and evaluating the impact of non-functional needs on operational requirements. Non-functional needs may influence decision-making when functional requirements are refined, resulting in the discovery of additional relevant requirements that should have been included in the original goal model. The SysML/KAOS model aids in establishing a hierarchy of needs in the form of objectives. The first stage requires that functional and non-functional needs be specified separately in a distinct goal model, but both at the same degree of abstraction. As a result, the ultimate goal model is developed.

### 4.4. RELAX

RELAX is considered a requirement engineering language for Dynamic Adaptive System (DAS) to handle uncertainty. Relax is expressed as a natural language that constitutes operators directed at finding fate. The types of uncertainty that could be addressed by RELAX, which relate to the amendments needed for the Dynamic Adaptive Systems to handle the kind of changes that cannot be taken due to its relative complexity [28][29]. These types could be:

- Environmental uncertainty: is when there is an unexpected change in the environmental conditions, such as noisy networks, random user input, failures of sensors, and malicious threats [30]. These situations lead to difficulties in maintaining the exact defined requirements in a different or unknown context.

- Behavioral uncertainty: is the case where the requirements need to amend. For instance, "the requirements of a space probe may change mid-flight to pursue science opportunities not foreseen by the designers". This type of uncertainty could define the requirements to change systems behavior during run time and reflect on environmental fate.

In RELAX, the focus is mainly directed on structured natural language requirements. Generally, the textual requirements define behavioral using modal verbs such as WILL (or SHALL), which shows the software system SHOULD provide [30]. In self-adaptive systems, environmental uncertainty could lead to difficulties in applying all the SHALL statements. Making a balance between the SHALL statements is important to relax non-critical statements for the more critical ones.

The RELAX operators help requirements engineers define the requirements that should not be tampered with and describe the requirements that could temporarily relax in some conditions. It could also help in determining the constraints on how to lead in making these requirements relax.

## 5. Evaluation and Discussion of Self-adaptive Systems

However, we were able to enhance the connection between critical ideas via the use of comparison criteria. We have been able to establish relationships between them. Table (2) shows a brief evaluation of the previously explained models. The assessment is based on four concepts (requirements description, monitoring, relationship, dependency/impact, and tools). While not all concepts are addressed entirely, we have highlighted in Table (2) the closest mechanism that supports the concepts. Requirements are represented as goals in SysML/KAOS; they are specified in the textual form in SYSML; they are defined in textual form in RELAX. This divides requirements into invariant and RELAX-ed conditions, each with its uncertainty factor. AND/OR relationships and contribution concepts are supported in SysML/KAOS. ContributionNature and ContributionType are two properties that describe the contribution's qualities. The first column indicates whether the contribution is positive or negative, and the second column indicates whether the help is direct or indirect. These connections are referred to as verify>> and refine>> in SysML. While REL establishes a link between ENV and MON, RELAX establishes a relationship between ENV and MON. SysML/KAOS defines Dependency/Impact as the impact of a non-functional goal (NFG) on a functional goal (FG); this effect may be positive or negative. The SysML/KAOS meta-Impact model's class (see [11]) is an association class between a Contribution Goal and an FG. It captures the fact that a Contribution Goal affects a Financial Goal, which is expressed as the Impact of a Non-Financial Goal on a Financial Goal. While SYSML has the derive>> concept, RELAX incorporates both positive and negative dependency. To manage to monitor, SysML/KAOS includes the Contribution objective concept used to complete an NFG. Additionally, SYSML contains the satisfy>> keyword, which indicates if a block>> satisfies a requirement>>. At the same time, RELAX makes use of the concept of MON to quantify the environment, i.e., ENV. The SysML/KAOS editor is a component of SysML/KAOS. Additionally, SYSML includes various additional tools, including eclipse, papyrus, top cased, and RELAX. We have the COOL RELAX editor, which is an eclipse-based application.

**Table (2): Evaluation of self-adaptive systems**

| Concepts/Approaches | SysML/KAOS | SysML | RELAX |
|---|---|---|---|

| Requirements Description | Abstract Goal Elementary Goal | Textual Requirement | Relaxed Requirement |
|---|---|---|---|
| Monitoring | Contribution Goal | <<satisfy>> | MON |
| Relationship | AND/OR | <<verify>> <<refine>> | REL |
| Dependency/Impact | Contribution nature Positive/Negative | <<derive>> <<contain>> | DEP: Positive Negative |
| Tools | Eclipse-based SysML/KAOS Editor | Eclipse | Eclipse-based COOL Relax Editor |

The KAOS principles are incorporated into the SysML/KAOS methodology, as shown in the preceding table. We have discovered that important modeling concepts perform better at the software level than at the requirements level in terms of comparison criteria.

## 6. Conclusion

This survey summarized the ideas, techniques, and difficulties associated with the emerging area of self-adaptive software systems straightforwardly and methodically. Thorough knowledge of this novel and the difficult subject will undoubtedly aid in implementing and promoting these concepts within the software engineering community. The fundamental part of this paper is defining the new classification of requirements that runtime behavior of requirement will take the advantages from it. Existing requirements and monitoring frameworks are two types of prototyping that are directed toward the expression of requirements. The actual usability of this requirement class for the adaptive system comes from the fulfillment of the looping feedback that yields adaptively, almost probably abstract the well known as plan-execute-monitor-analyze the loop feedback researching the computing that comes from autonomic. The application of the feedback loop structure defines a scientific technique for adaptive system analysis based on integrating and providing awareness requirement for such construction in frameworks, regarding the awareness requirement itself, improvement of definition in formalization of awareness requirement, and consolidation of the model of domain model within the approach which will support the initiating adaptively of awareness requirement. This work has attempted to establish the groundwork for self-adaptive software engineering as a mature discipline capable of using current systems and not relying only on technical advancements for advancement.

## Acknowledgments

The authors would like to express their heartfelt appreciation to the University of Kurdistan Hewler for providing the facilities and ongoing assistance necessary to perform this research.

## Ethical Standards

**Conflict of Interest:** The authors state that they are not involved in any conflict of interest.

**Funding:** Funding is not applicable.